\documentclass[conference]{IEEEtran}
\IEEEoverridecommandlockouts
\usepackage{cite}
\usepackage{amsmath,amssymb,amsfonts}
\usepackage{algorithmic}
\usepackage{graphicx}
\usepackage{textcomp}
\usepackage{xcolor}
\usepackage{times}
\usepackage{caption}
\usepackage{booktabs}
\usepackage{makecell}
\usepackage{multirow}
\usepackage{caption}
\usepackage{subcaption}
\usepackage{hyperref}
\usepackage{setspace}
\def\BibTeX{{\rm B\kern-.05em{\sc i\kern-.025em b}\kern-.08em
    T\kern-.1667em\lower.7ex\hbox{E}\kern-.125emX}}

\makeatletter
\def\@IEEEpubidpullup{8\baselineskip}
\makeatother

\begin{document}

\IEEEoverridecommandlockouts
\IEEEpubid{
\parbox{\columnwidth}{\vspace{-4\baselineskip}Permission to make digital or hard copies of all or part of this work for personal or classroom use is granted without fee provided that copies are not made or distributed for profit or commercial advantage and that copies bear this notice and the full citation on the first page. Copyrights for components of this work owned by others than ACM must be honored. Abstracting with credit is permitted. To copy otherwise, or republish, to post on servers or to redistribute to lists, requires prior specific permission and/or a fee. Request permissions from \href{mailto:permissions@acm.org}{permissions@acm.org}.\hfill\vspace{-0.8\baselineskip}\\
\begin{spacing}{1.2}
\small\textit{ASONAM '19}, August 27-30, 2019, Vancouver, Canada \\ 
\copyright\space2019 Association for Computing Machinery. \\
ACM ISBN 978-1-4503-6868-1/XX/XX?/\$XX.XX \\
\url{http://dx.doi.org/XX.XXXX/XXXXXXX.XXXXXXX}
\end{spacing}
\hfill}
\hspace{0.9\columnsep}\makebox[\columnwidth]{\hfill}}
\IEEEpubidadjcol

\title{Uncovering Download Fraud Activities in Mobile App Markets}

\author{\IEEEauthorblockN{Yingtong Dou\IEEEauthorrefmark{1},
Weijian Li\IEEEauthorrefmark{2},
Zhirong Liu\IEEEauthorrefmark{3}, 
Zhenhua Dong\IEEEauthorrefmark{3},
Jiebo Luo\IEEEauthorrefmark{2} and Philip S. Yu\IEEEauthorrefmark{1}\IEEEauthorrefmark{4}}
\IEEEauthorblockA{\IEEEauthorrefmark{1}Department of Computer Science, University of Illinois at Chicago, USA}
\IEEEauthorblockA{\IEEEauthorrefmark{2}Department of Computer Science, University of Rochester, USA}
\IEEEauthorblockA{\IEEEauthorrefmark{3}Noah's Ark Research Lab, Huawei, China}
\IEEEauthorblockA{\IEEEauthorrefmark{4}Institute for Data Science, Tsinghua University, China\\ \{ydou5, psyu\}@uic.edu, \{wli69, jluo\}@cs.rochester.edu, \{liuzhirong, dongzhenhua\}@huawei.com}}

\maketitle

\begin{abstract}
Download fraud is a prevalent threat in mobile App markets, where fraudsters manipulate the number of downloads of Apps via various cheating approaches. Purchased fake downloads can mislead recommendation and search algorithms and further lead to bad user experience in App markets. In this paper, we investigate download fraud problem based on a company's App Market, which is one of the most popular Android App markets. We release a honeypot App on the App Market and purchase fake downloads from fraudster agents to track fraud activities in the wild. Based on our interaction with the fraudsters, we categorize download fraud activities into three types according to their intentions: boosting front end downloads, optimizing App search ranking, and enhancing user acquisition\&retention rate. For the download fraud aimed at optimizing App search ranking, we select, evaluate, and validate several features in identifying fake downloads based on billions of download data. To get a comprehensive understanding of download fraud, we further gather stances of App marketers, fraudster agencies, and market operators on download fraud. The followed analysis and suggestions shed light on the ways to mitigate download fraud in App markets and other social platforms. To the best of our knowledge, this is the first work that investigates the download fraud problem in mobile App markets.
\end{abstract}


\section{Introduction}
Nowadays, smart devices enable people to access various services through mobile applications (Apps) easily. Apps are distributed by the App markets, which are digital platforms developed by third-party Internet companies, e.g., the Google Play by Google, or smart device vendors such as the iOS App Store by Apple. In App markets, users can browse Apps with various types of contents and functions, while the App markets make profits through distributing the paid Apps and advertisements. According to a recent study \cite{martin2017survey}, App markets now cover more than 2 billion devices and are worth more than \$25 billion over the rapid growth all over the world.  

The considerable profit leads to numerous fraud activities targeting the factors that determine the App choices of users. Previous studies have investigated ranking fraud \cite{rahman2017search, zhu2015discovery, chen2017toward}, spam reviews \cite{seneviratne2015early, li2017crowdsourced}, App removal mechanism \cite{wang2018android} and malware dissemination \cite{chen2015finding, rahman2016fairplay} in App markets.

Among all fraud activities in App markets, download fraud aims at manipulates the quantity of App downloads. Evidences from news \cite{fakedownloads} and black market \cite{blackmarket} indicate that download fraud is inundant in App markets. A recent report from Datavisor \cite{datavisor} claims that near 10\% downloads\&installs in mobile marketing are fake, which causes up to \$300 million loss for app marketers in 2018. For App market itself, the injected fake downloads would mislead the recommender system to generate low-quality recommendation results or even disturb the whole App market ecosystem. 

In order to obtain a holistic view of download fraud activities in App markets, we investigate them from multiple perspectives. Different from previous works \cite{rahman2017search, zhu2015discovery, chen2017toward, seneviratne2015early, li2017crowdsourced, wang2018android, chen2015finding, rahman2016fairplay}, where the data are crawled from App market portals, which lack fraudsters information and ground truth of fraud activities, we dive deeper and explore the download fraud activities inside a company's App Market, which is one of the most popular Android App markets in the world \cite{wang2018beyond}. With access to the billions of server-side download data, we aim to answer the following research questions:

\begin{itemize}
\item \textbf{RQ1:} \textit{What are the types of download fraud activities in the App market?}
\item \textbf{RQ2:} \textit{How to identify the download fraud activities?}
\item \textbf{RQ3:} \textit{How to mitigate the download fraud in App markets?}
\end{itemize}

To answer RQ1, we set up a honeypot App and disguise as an App marketer to purchase fake downloads from fraudster agencies. Integrating the information from the server-side download log and agencies, we find three major types of download fraud activities in App markets with different goals. 

For RQ2, we find a solid method to acquire the ground truth of download fraud activities. Then we design a number of features and train machine learning models to identify fake downloads and suspicious Apps involved in download fraud campaigns. 

For the third research question, we interview three distinct parties (App marketers, fraudster agencies, and market operators) related to App download fraud activities. Based on their stances and our analysis, we offer several guidance and suggestions for preventing the download fraud from the perspectives across security, App market operation, and advertisement.

In summary, our work makes three major contributions:

\begin{itemize}
\item We investigate and categorize the download fraud activities in three different types under an industrial setting. To the best of our knowledge, it is the first work to conduct comprehensive investigations on download fraud in mobile App markets. (Section \ref{sec:rq1}.)
\item We propose new features as well as adapting features from previous works in identifying fake downloads. The experiment on a large scale industrial dataset extracts a number of informative features and discover meaningful patterns of fraud activities. (Section \ref{sec:rq2}.)
\item We interview and present the stances of three parties involved in download fraud activities, as well as our suggestions on mitigating download fraud and building a better App market ecosystem. (Section \ref{sec:rq3}.)
\end{itemize}

\section{Related Work}
\label{sec:rw}

Our work lies on three major research topics: App markets security, click fraud detection, and black market investigation.

Previous works have investigated various kinds of security issues in App markets. Chen \cite{chen2015finding} and Rahman \cite{rahman2016fairplay} analyzed malware dissemination in Google Play. Zhu \cite{zhu2015discovery} and Chen \cite{chen2017toward} studied the suspicious Apps involved in search ranking in iOS App Store. Li et al. \cite{li2017crowdsourced} delved crowdsourced spam reviews in both Google Play and iOS App Store. Potharaju et al. \cite{carbunar2015longitudinal} gave a longitudinal analysis of Apps in Google Play and provided suggestions on detecting search ranking fraud. According to \cite{Appmasters}, Google Play does not eliminate all fake downloads. Moreover, few previous works have investigated this problem either. It is mainly due to the lack of the ground truth of fraud activities. The data crawled from the front end, which has limited information, also hinders previous work for a comprehensive study on the download fraud. In this work, with server-side data and device vendor information as the ground truth, we could take a holistic approach to probe the download fraud in App market.

The outcome of download fraud is similar to click fraud, which is a type of fraud that occurs in pay-per-click online advertising \cite{pearce2014characterizing}. Click fraudsters usually inject fake clicks to target URLs using click bots and steal money from advertisers. To detect click fraud, Pearce et al.  \cite{pearce2014characterizing} employed peer-to-peer measurements, command-and-control telemetry, and contemporaneous click data to analyze click fraud on botnets. Oentaryo et al. \cite{oentaryo2014detecting} devised various temporal and statistical patterns to detect click fraud in online advertising. Cao et al. \cite{cao2014behavioral} leveraged behavior features and click patterns to detect spam URL sharing. The download fraud we investigated in this paper is more complicated than click fraud (i.e., mixed with human and bot activities). Inspired by the click fraud detection works mentioned above, we propose to model the download fraud activities in a multiview and feature-based perspective.

For the black markets investigation, several works \cite{wang2012serf, lee2014dark, choi2016detecting, kaghazgaran2018combating} have probed the crowdsourcing websites and devised machine learning approaches to detect crowdturfing campaigns and crowd workers. Other works like \cite{xie2015Appwatcher} inspected the transactions over trading App reviews and \cite{tian2015crowd} investigated the crowd fraud in Internet advertisement. However, seldom previous work has studied the black markets targeting download fraud. Like \cite{lee2010uncovering,de2014paying, onaolapo2016happens}, we launch a honeypot App in App market to acquire reliable ground truth of download fraud activities. Moreover, we infiltrate into the black market and reap useful information from fraudsters to help our analysis.

\begin{figure}
\centering
\includegraphics[height=2in, width=3.45in]{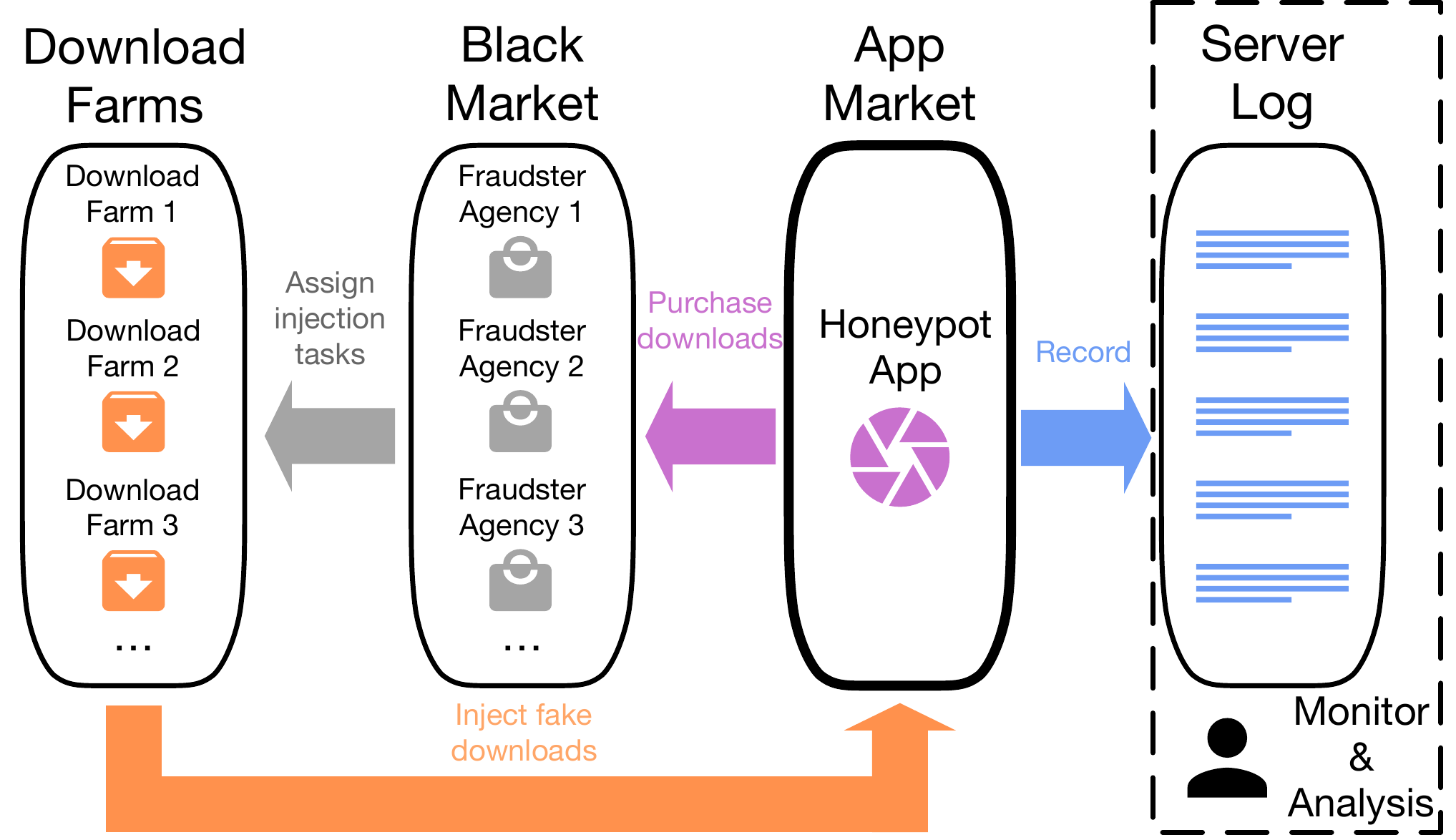}
\caption{The workflow of how we set up the honeypot and track fraud activities. We disguise as an App marketer and purchase fake downloads from fraudster agencies who plan and distribute downloads injection tasks to different download farms. The assigned farms inject fake downloads to our honeypot App in the App market. We monitor and analyze the fraud activities from the server-side log.}
\label{fig2}
\end{figure}

\begin{table*}
  \centering
  \caption{Summary of three download fraud types. They have different goals and employ different attacking approaches.}
  \label{tab1}
  \resizebox{\linewidth}{!}{%
  \begin{tabular}{ccccccc}
    \toprule
    \textbf{Fraud Type}&\textbf{Goal}&\textbf{Target}&\textbf{Approach}&\makecell{\textbf{Price}\\(USD/10k)}&\makecell{\textbf{Level}\\\textbf{of Threat}}&\makecell{\textbf{Detection}\\\textbf{Difficulty}}\\
    \midrule
    1 & Boosting front end downloads & Front end & Automated scripts & 5 & Low & Easy\\
    2 & Optimizing App search ranking & Back end & Download bots & 70 & High & Hard\\
    3 &  Enhancing user acquisition \&  retention rate & App & Crowd workers & 1400 & Low & Very hard\\
    \bottomrule
  \end{tabular}}
\end{table*}

\begin{table*}
  \centering
  \caption{Comparison between purchased fake downloads injection services on our honeypot App. Portal website: download comes from App market portal website. Update: download comes from updating the App. Null: no download source record.}
  \label{tab2}
  \resizebox{\linewidth}{!}{%
  \begin{tabular}{ccccccccc}
    \toprule
    \textbf{Farm Name}&\textbf{Access via}&\textbf{\#Downloads}&\textbf{Source}&\textbf{Price}(USD/10k)&\textbf{IP Address}&\textbf{Device ID}&\textbf{Duration}(hours)&\textbf{Date}\\
    \midrule
    Farm 1 & Website & 10,000 & Portal site & 4 & Distinct & None & 12 & 06/06/2018\\
    Farm 2 & Taobao & 15,000 & Update & 6 & Distinct & Normal & 2 & 07/31/2018\\
    Farm 3 & QQ & 10,000 & Null & 6 & Distinct & Abnormal & 0.2 & 08/05/2018\\
    Farm 4 & Website & 20,000 & Portal site & 3 & Distinct & Abnormal & 1 & 09/15/2018\\
    \bottomrule
  \end{tabular}}
\end{table*}

\section{RQ1: Download Fraud Types}
\label{sec:rq1}

We set up a honeypot App and infiltrate into the black market to explore the fine-grained download fraud types. Some terminologies used throughout this paper are summarized as follows.

\begin{itemize}
\item \textbf{Fraudster Agency} is a proxy in black market that connects download farms with their customers (i.e., app marketers and developers). It is usually in the form of an App promotion website.
\item \textbf{Download Farm} (also known as Click Farm) is a physical location that conducts fake download injection service. Download farm is usually composed of numerous real mobile devices or device simulators, we call them the download bots \cite{datavisor}.
\item \textbf{App Store Optimization (ASO)} is a process that applies a set of optimization policies to Apps in order to boost App search ranking in App market.
\item \textbf{Crowdturfing} is the campaign initialized by App marketers. Crowd workers obtain monetary rewards in exchange for performing simple \textquotedblleft tasks" inside Apps that go against accepted user policies \cite{wang2012serf}.
\end{itemize}

\subsection{Setting Up the Honeypot}

Suspicious Apps involved in massive download fraud are easy to detect when we examine their download traffics. However, it is difficult to identify whether each download of an App is fake or not, since fake downloads usually mix with the legitimate ones. Inspired by setting honeypots in Twitter \cite{lee2010uncovering} and Facebook \cite{de2014paying} to acquire the ground truth of fraud activities, we also place a standalone honeypot App \footnote{The released honeypot App is a gaming App which is the most frequently purchased App according to fraudster agencies.} in the company's App market. We track the activities of download fraudsters from the server-side log, and further dissect their working mechanism along with the information probed from fraudsters.

Figure \ref{fig2} illustrates our honeypot setting workflow. Since most of the fraudsters are in the black market, we connect with the fraudster agencies found by searching keywords such as \textit{App promotion}, \textit{App store optimization} and \textit{buy App downloads} on search engines. As Table \ref{tab2} shows, we finally contact with fraudster agencies via their websites, e-commerce platforms like Taobao, or online chatting services like QQ. To make sure the honeypot App does not influence the App market regular operation, we set the honeypot App to be inaccessible by users in the App market. Therefore, the server log only has the records of purchased fake downloads. Meanwhile, we are able to monitor various attributes of download fraud activities given access to the server-side data.

After setting up the honeypot, we purchase four download injection services where each of them is asked to inject 10k-20k fake downloads to our honeypot App separately. We select four distinct periods among four months in 2018 to avoid potential bias.

\subsection{Fraud Types}

During communicating with fraudster agencies, we observe various kinds of fake downloads injection service. Based on this, we try to categorize those frauds under a unified scheme and expect a full picture of download fraud activities. Integrating the information from fraudster agencies and our server-side observation, we classify download fraud activities into three major types according to their goals. Table \ref{tab1} summarizes their characteristics.

\subsubsection{\textbf{Boosting Front End Downloads}}

This type of fraud targets at increasing specific App's download times displayed at the App market web  portal and mobile client (a.k.a. front end), which could be tracked by third-party App analytics companies. The displayed download count is a significant indicator for App analytics companies to provide App quality assessments. Thus, a boost in the front end downloads display could help the App gain a better rating. According to fraudster agencies, such fraud is relatively an easy task that could be implemented by automated scripts. Download farms deploy similar injecting techniques in injecting this type of downloads. Such fake downloads are at low cost compared to other two types of fake downloads and only take effect at the front end.

From the server-side log, we find that all purchased fake downloads to the honeypot App fall into this category. Table \ref{tab2} shows the main attributes of four purchased download injection services. We could see that fake downloads injected by four farms all have distinct IP addresses but vary in downloading source and device ID. Regular downloads usually come from App market clients on smart devices, while Farm 1 and Farm 4 both inject fake downloads from the App market portal website. The mechanism is simply clicking the honeypot App downloading URL listed at the App market portal website. Moreover, they both fail to generate valid device IDs. Farm 2 can simulate the device ID information, but it injects the fake downloads via repeatedly updating the honeypot App, which is not an ordinary behavior.

\subsubsection{\textbf{Optimizing App Search Ranking}} 

In App markets, users' search results or App recommendation results will be displayed as lists of ordered Apps to the users. Apps are ranked according to an elaborate algorithm where the amount of App downloads is a crucial feature. The Apps listed at the top of the list are supposed to be the most related target, and have a higher chance to be viewed and further downloaded by users. Thus, boosting an App's ranking order could increase App discovery and organic downloads. However, according to our investigation, the first type of download fraud targeting the front end display has little effect on the company's App ranking system. To mislead the App ranking system and thereby increase the app exposure rate, the second type of download fraud injects more \textquotedblleft genuine" downloads to server logs. Specifically, download farms employ download bots with more advanced attacking techniques to inject fake downloads. Those camouflaged downloads are usually considered as genuine downloads. 

From the information probed from fraudster agencies, injecting legitimate-like fake downloads is usually a part of App Store Optimization service, and it would cost at least 300 US Dollars for our honeypot App. Purchasing such ASO service to our honeypot App incurs a large budget and is a time-consuming process. Meanwhile, we are informed that all injected downloads come from download bots, which are not real devices and may contain irregular device vendor flags. To examine the hypothesis, we make use of the device vendor flag and experiment with server-side logs. More details are discussed in Section \ref{sec:rq2}.

\begin{figure}[!h]
     \centering
     \begin{subfigure}[b]{0.23\textwidth}
         \centering
         \includegraphics[width=\textwidth]{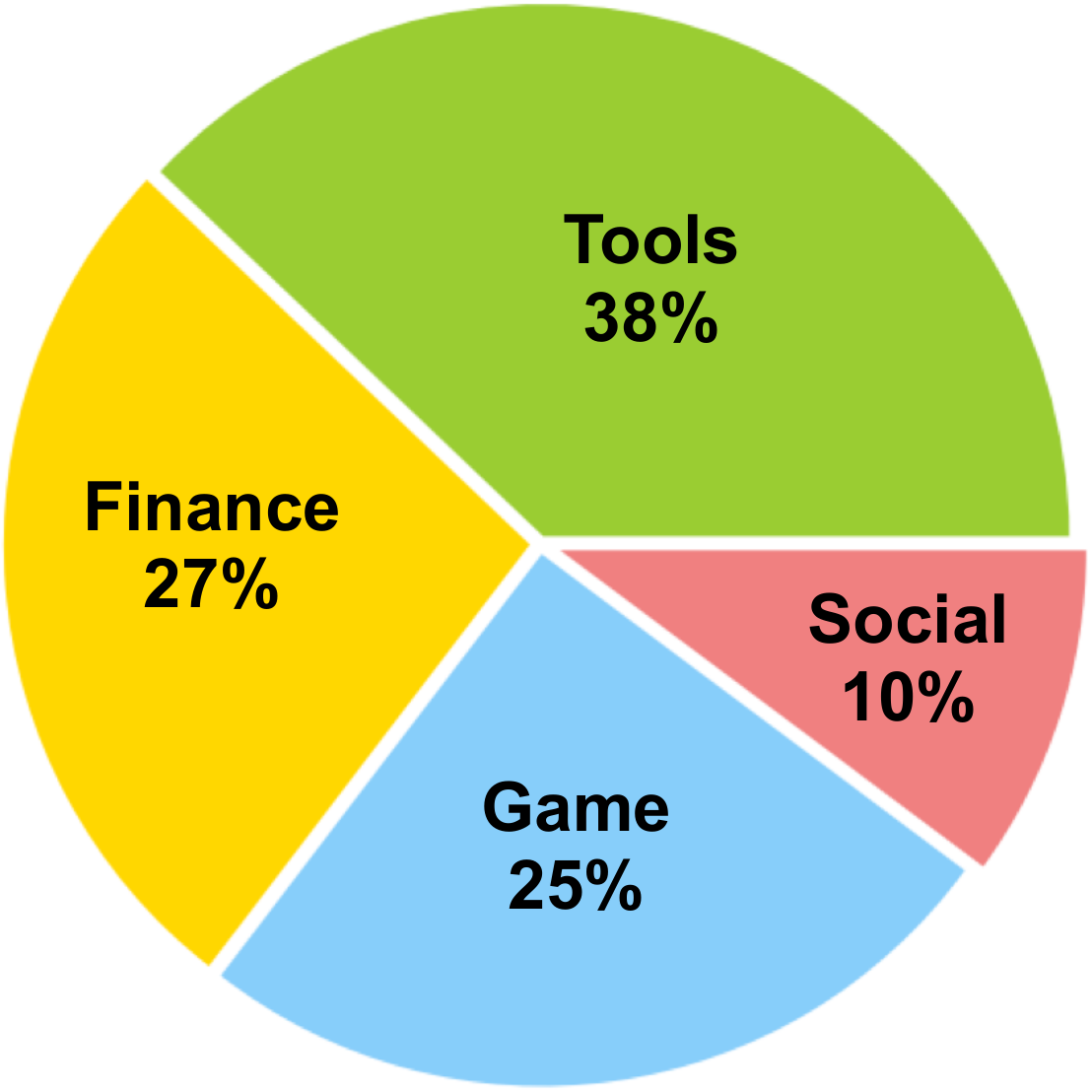}
         \caption{}
         \label{fig3a}
     \end{subfigure}
     \hfill
     \begin{subfigure}[b]{0.23\textwidth}
         \centering
         \includegraphics[width=\textwidth]{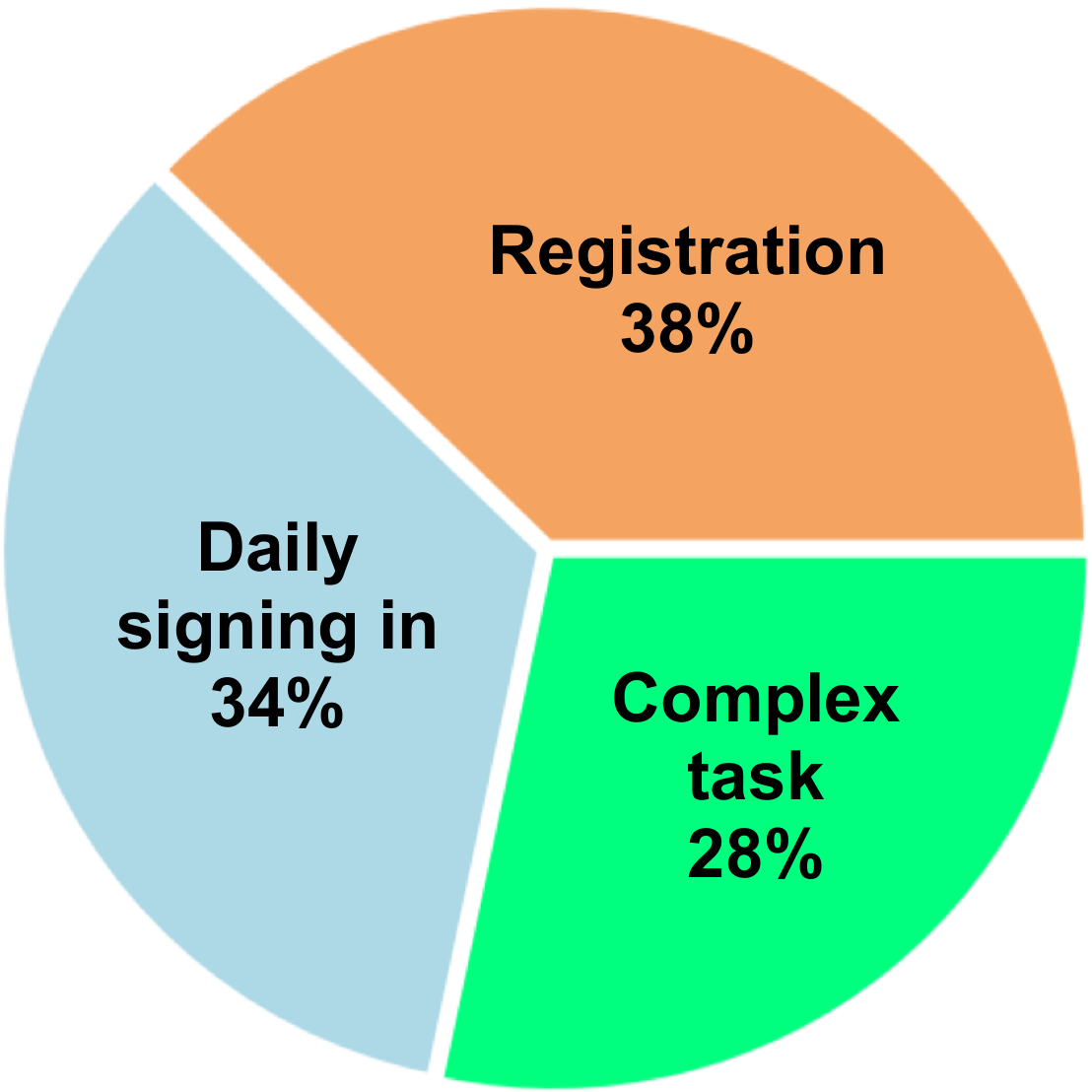}
         \caption{}
         \label{fig3b}
     \end{subfigure}
\caption{The distribution of (a) categories of Apps that solicit crowdturfing services; (b) post-install crowd tasks requiring user engagement.}
\label{fig3}
\end{figure}

\subsubsection{\textbf{Enhancing User Acquisition \& Retention Rate}}

While the first two types of download fraud aim at cheating the App market with fake App downloads, the last type is of large difference, which focuses on the App itself. Its goal is to increase or keep user acquisition \& retention rate mostly for business purposes, such as achieving the product KPIs to attract investments. This type of fraud is more complex and adopts crowdturfing attack to manipulate target Apps. 

The crowdsourcing tasks are commonly published on crowdsourcing websites (a.k.a. App trial platforms). Any user with a smart device can register as a crowd worker on the website and seek the task they would like to do. After the required task is done, the platform would pay crowd workers monetary reward. Since most fraud activities of this type are executed inside the Apps, and it produces only a few amounts of new downloads, its threat to App markets is much lower than the first two types of fraud. From App market server-side log, such download fraud activities are extremely difficult to track, since the behavior of crowd workers inside App market is almost the same as regular users.

To get a sense of this type of fraud, we crawl information from sixteen App trial platforms in China. Figure \ref{fig3} summarizes the collected information. Figure \ref{fig3a} indicates that more than half of the plagiarism Apps are Finance and Game Apps because they can monetize users. The Apps in Tools usually integrate with many mobile advertisements, which makes Tools become the leading category purchasing such fraud service. According to Figure \ref{fig3b}, Registration and Daily signing in correspond to user acquisition and retention. The post-install tasks like reposting news, adding bank accounts, and playing games are personalized requirement from App marketers, which are related to KPIs. The price of the crowdturfing task depends on its complexity. In general, its average cost is much higher than the first two types of download fraud.

\section{RQ2: Identifying Fake Downloads}
\label{sec:rq2}

Setting up honeypot helps us capture the evidence of the first type of download fraud activity, and further figure out the multiple facets behind the fake downloads. According to our analysis in Section \ref{sec:rq1}:

\begin{itemize}
\item The records of the fake downloads injected by the first type of download fraud have distinct differences from other download records. Those fake downloads could be easily determined and filtered with the Source and Device ID information, as shown in Table \ref{tab2}.
\item The third type of download fraud involves multiple factors across the App markets. We will discuss the approaches to mitigate it along with the stances from different parties in App market ecosystem in Section \ref{sec:rq3}.
\end{itemize}

For the second type of fraud, the information we obtained from the download agencies and the third-party reports indicates that it is the most prevalent download fraud type in mobile App markets \cite{Appmasters, chen2017toward, wang2018beyond}. However, previous works are unable to address the problem from a holistic view due to the data limitation and lack of ground truth \cite{ zhu2015discovery, chen2017toward}. For this reason, with the access to server-side data, we attempt to discover useful features that could spot the bot-generated downloads, and give an in-depth analysis of the second type of download fraud in this section.

\subsection{Ground Truth \& Data Collection}

According to the fraudster agencies, the second type of download fraud is executed by bots controlled by download farms. However, those bots are not able to simulate device IDs produced by smartphone vendors. Also knowing that more than 90\% of total App downloads in the company's App Market are from smartphones produced by the company, a download record from a non-vendor device is very suspicious. Therefore, we could utilize the vendor flag in the download logs as the ground truth to help us identify the bot activities. 

We define the \textbf{\textit{negative download record}} (\textbf{\textit{normal download}}) as the download record of vendor-verified device. We consider App with more than 50\% of downloads from non-vendor devices during the experiment period as the \textbf{\textit{suspicious App}} involved in download fraud activities. The \textbf{\textit{normal Apps}} are Apps whose downloads are all originated from vendor-verified devices. Then, all download records of the suspicious Apps during the experiment period are regarded as \textbf{\textit{positive download records}} (\textbf{\textit{suspicious downloads}}). By doing so, we could ensure that positive samples contain bot records and extensive data for a valid study.

By sampling over billions of server-side download records in ten random days among six months, we could conduct an accurate quantitative analysis without potential bias. Our final experiment dataset includes around one million positive records and nine million negative records, which covers more than half a million Apps. We strictly obey the privacy protection policies when collecting the data. All device IDs and IP addresses are encrypted with hash codes. Therefore, the dataset contains no personal information and is completely anonymous.

\begin{figure}
\includegraphics[height=3.35in, width=3.4in]{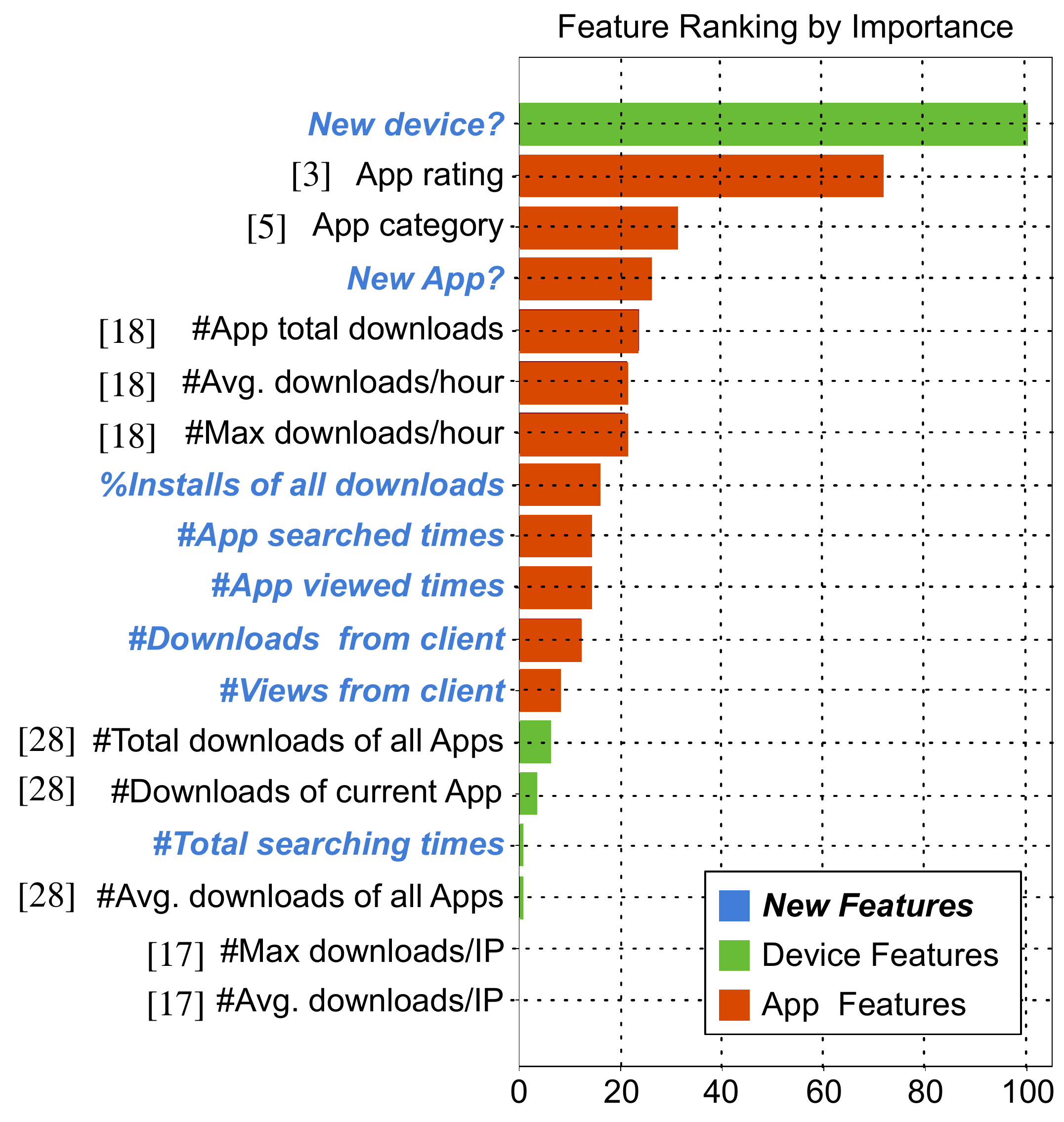}
\caption{Selected features of a unique download record from server log. The relative importance is calculated according to Gini Index. \textbf{New device?} or \textbf{New App?} represents whether the App or device is released or first activated within one week of the record timestamp.} 
\label{fig4}
\end{figure}

\subsection{Feature Selection \& Validation}
\label{feature}

With the ground truth data, we aim to discover features that could identify fake downloads efficiently. Moreover, we would like to validate the effectiveness of features proposed by the state-of-the-art works. To facilitate future analysis under the industrial environment, the selected features should be interpretable as well. According to the intuition above, we decide to design features based on server-side data, previous works, and information from fraudster agencies. 

Each record in the download log has the metadata of one download action (e.g., device ID, App ID, timestamp). We could query more metadata and statistical metrics of devices and Apps from other server-side logs via device ID and App ID. According to previous work modeling behavior anomalies \cite{cao2014behavioral, rayana2015collective}, we harness the statistical metrics such as \textbf{\textit{\#Avg.\&Total downloads of all Apps}} to model general download behavior and \textbf{\textit{\#Avg.\&Max. downloads/hour}} to model the burst in download traffics. Besides behavior features, we also select App metadata, like \textbf{\textit{App category}}, and \textbf{\textit{App rating}} which probably can distinguish suspicious Apps from regular ones \cite{zhu2015discovery, seneviratne2015early}. According to prior works on click fraud detection \cite{oentaryo2014detecting}, bots inside the same group are supposed to share the same IP address. Thus, we use \textbf{\textit{\#Max.\&Avg. downloads/IP}} to model it.

From the download farms, we are informed that the second type of download fraud usually targets new Apps and could simulate regular user behavior like searching and viewing Apps before hitting the download button. Along with other server-side data which is supposed to be as signals of bot activities (search, view, download from clients), we add \textbf{\textit{New Features}} shown in Figure \ref{fig4}.

To evaluate the capability of selected features in identifying fake downloads, we use Gini Impurity to calculate feature importance \cite{geurts2006extremely}. Figure \ref{fig4} shows the normalized relative importance of the selected features. From the feature importance ranking, we have the following observations and conclusions:

\begin{itemize}

\item The most informative feature \textit{New device?} indicates that download bots usually reset their device IDs after one download action. Because it is difficult to determine the suspiciousness of a new device with no history record.

\item App rating and App category are two other top informative features. It reveals that the Apps involved in download fraud activities are different in attributes from regular Apps. More details are discussed in Section \ref{sec4.3}.

\item Many Apps involved with download fraud are new released Apps. It reflects the intention of App marketers to purchase fake downloads to facilitate their App launching.

\item Except for the first device feature, most of the App download statistics features reveal more signals than device behavior features in identifying fake downloads.

\item App statistics like installations, views, searched times, and client downloads help distinguish the abnormal traffics. It validates our assumptions when selecting those features.

\item Most of the device behavioral features and IP-based features have little contribution to the classification task. It contradicts to our early assumptions, showing that the download bots can indeed simulate regular users' behavior very well, which is similar to the observation on social spambots in previous study \cite{cresci2017paradigm}.

\item The total searching times of bots are similar to regular users. It indicates that bots could emulate the searching behavior of regular users. Meanwhile, suspicious Apps have higher searched times than regular Apps. Both of them reflect that the fraudsters engaged in the second type of download fraud aim to manipulate the app search ranking in an imperceptible way.

\end{itemize}

To further validate the performance of selected features, we feed different types of features into XGBoost which is an efficient tree boosting based classifier \cite{chen2016xgboost} and test them on a separated validation dataset collected on January 2019. 

\begin{figure*}
     \centering
     \begin{subfigure}[b]{0.32\textwidth}
         \centering
         \includegraphics[width=\textwidth]{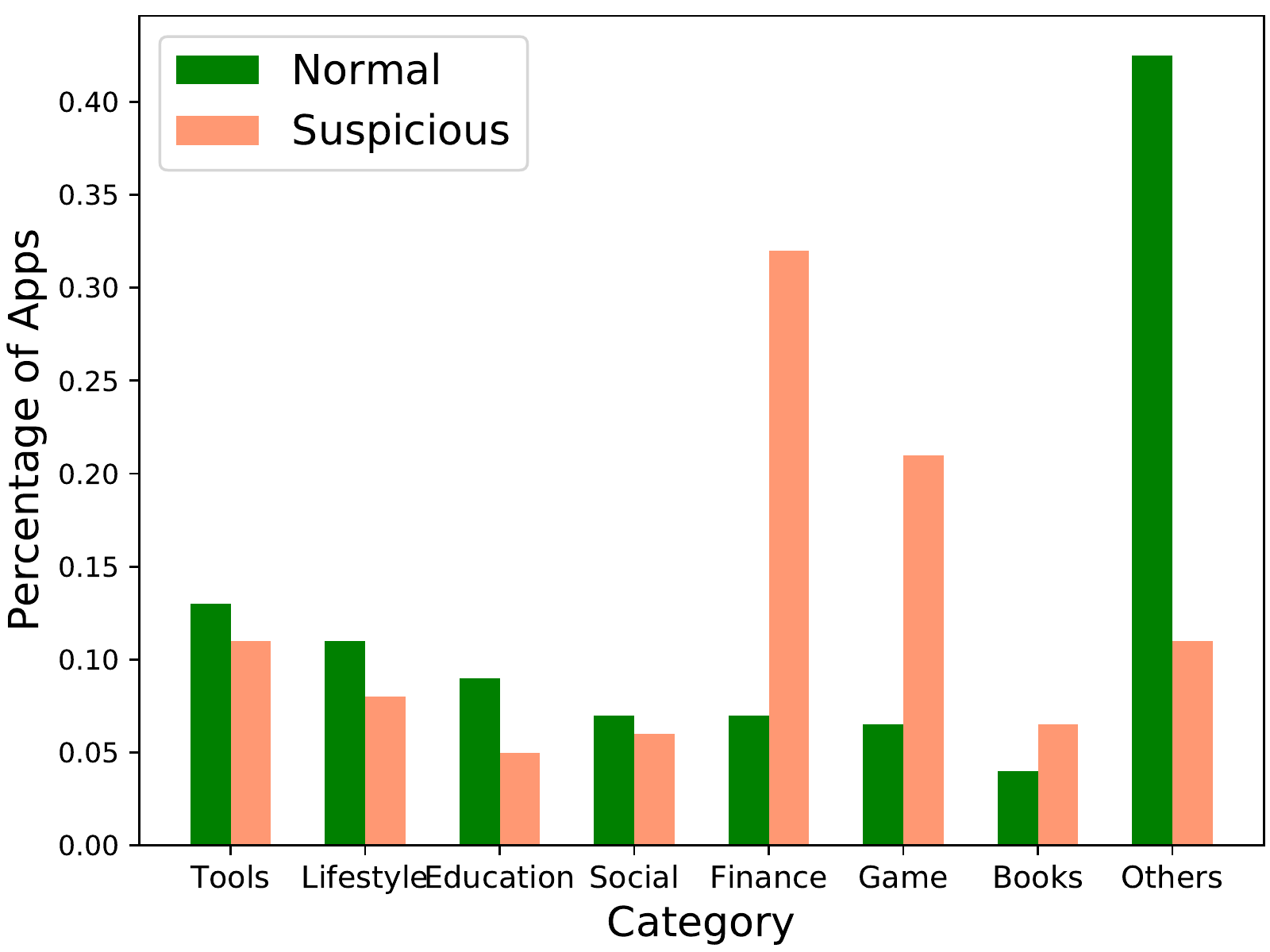}
         \caption{}
         \label{fig5}
     \end{subfigure}
     \hfill
     \begin{subfigure}[b]{0.32\textwidth}
         \centering
         \includegraphics[width=\textwidth]{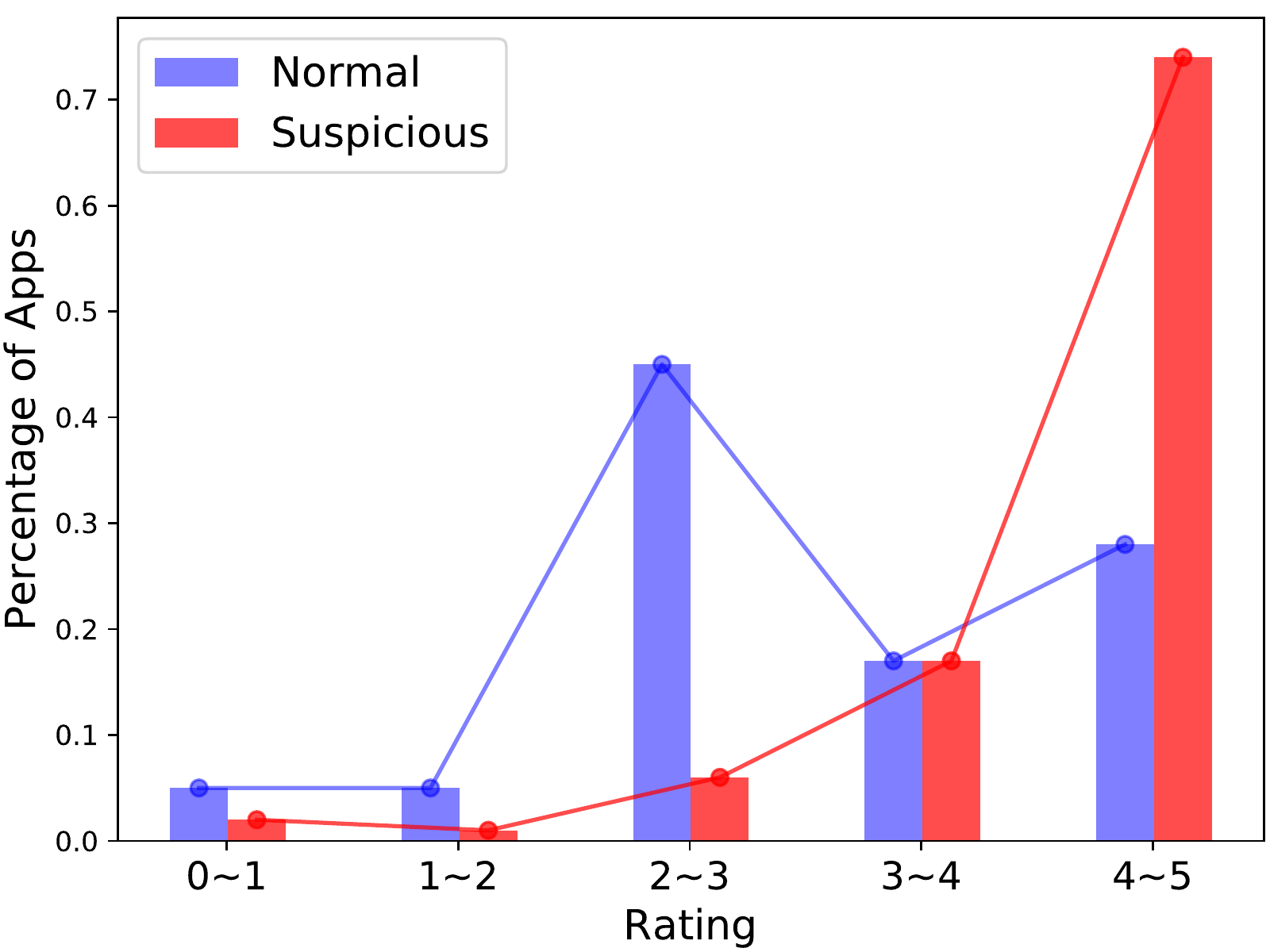}
         \caption{}
         \label{fig6}
     \end{subfigure}
     \hfill
     \begin{subfigure}[b]{0.32\textwidth}
         \centering
         \includegraphics[width=\textwidth]{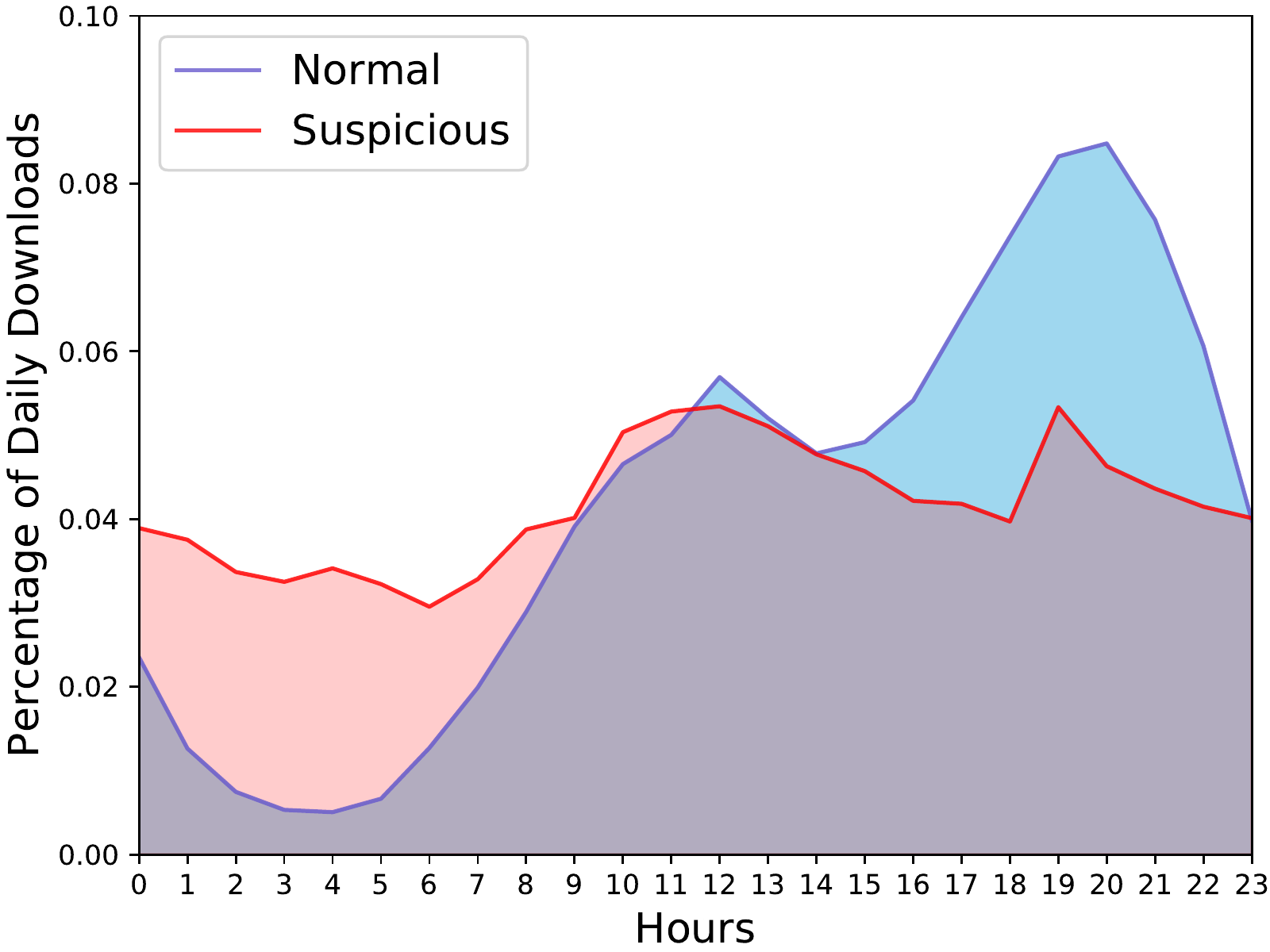}
         \caption{}
         \label{fig7}
     \end{subfigure}
        \caption{Characteristics of suspicious Apps and downloads comparing with normal Apps and downloads. (a) App category distribution of suspicious Apps versus normal Apps; (b) rating distribution of suspicious Apps versus normal Apps; (c) suspicious downloads traffic versus normal downloads traffic in a day.}
\end{figure*}

Table \ref{tab3} shows the testing result. We note that the device features set could identify more fake downloads, but it has a lower precision, which means it classifies more regular downloads as fake. A high false positive rate will make the legitimate app marketers compromise mobile App markets, and may further reduce the markets' revenue. Though the features proposed by us cannot beat the performance of previous features, aggregating all features together could hit 100\% among all metrics. It validates the rationale behind our feature selection.

\begin{table}[!h]
  \centering
  \caption{Testing results of different types of features with XGBoost.}
  \label{tab3}
  \resizebox{\linewidth}{!}{%
  \begin{tabular}{lccccc}
    \toprule
    \textbf{Feature Type}&\textbf{Precision}&\textbf{Recall}&\textbf{F1}&\textbf{AUC}&\textbf{Accuracy}\\
    \midrule
    Device & 0.955 & 0.988 & 0.963 & 0.977 & 0.992\\
    App & 0.978 & 0.972 & 0.976 & 0.965 & 0.993\\
    \midrule
    New & 0.974 & 0.940 & 0.951 & 0.969 & 0.993\\
    Previous & 0.974 & 0.977 & 0.975 & 0.996 &0.987\\
    \midrule
    All & \textbf{0.994} & \textbf{0.992} & \textbf{0.993} & \textbf{0.998} & \textbf{0.997}\\
    \bottomrule
  \end{tabular}}
\end{table}

\subsection{Comparative Analysis}
\label{sec4.3}

To gain a more in-depth insight into the second type of download fraud, we conduct a comparative analysis between fraud bots/Apps and regular users/Apps with the identified fake downloads and suspicious Apps.

With the XGBoost classifier above, we have identified more than one hundred suspicious Apps among all apps in the company's App Market from June 2018 to December 2018. Figure \ref{fig5} shows the category distribution of fraud Apps comparing with the overall category distribution among all Apps across the App market. Notably, Finance and Game Apps account for more than half of the suspicious Apps, while these two kinds of Apps only take up to 15\% among all Apps in the App market. We consider Game and Finance Apps have more potential in monetizing users, which makes a significant amount of profit. This profit leads to the interests in soliciting download fraud campaigns. The observation is consistent with information probed from fraudster agencies.

As a critical metric in App market, App rating also attracts our great interest. Fig \ref{fig6} shows the rating distribution of suspicious Apps comparing with normal Apps across the App market. Normal Apps and suspicious Apps have different rating distributions. The mode and mean of normal App ratings are both in the range 2-3 and 3-4, while the mode and mean value of suspicious Apps are in 4-5. A significant portion of the extreme positive ratings of suspicious Apps validates the conclusion of Zhu et al. on ranking fraud detection \cite{zhu2015discovery}. It demonstrates that download fraud usually comes with rating manipulation, and they aim to fool the App ranking algorithms via manipulating App downloads and ratings.

In the perspective of the download activities, we plot the normalized hourly download times of fake downloads and normal downloads in Figure \ref{fig7}. Intuitively, regular users have less activity during the resting time, so there is seldom downloads occur during this period. On the other hand, bots do not have resting time, and their download activities are relatively steady throughout the day. The different distribution further proves that fake downloads of the second fraud type are generated by bots.

There are two more interesting phenomenons we find while analyzing the suspicious Apps. The first one is that not all anomalies are fraudulent. We find some Apps with downloads burst during a short period are actual at their promotion phases. From Google Trends\footnote{\url{https://trends.google.com}}, the App searching trends are consistent with their downloads traffic. It indicates that the traffic spikes may be made by other promotion channels outside App markets. A similar phenomenon has been observed in user-review social networks as well \cite{zheng2017smoke}. Another observation is that the categories suffering from download fraud vary with time. For example, a large amount of sports betting Apps involved in download fraud are identified during the 2018 FIFA World Cup. While, in December 2018, there are only a few amounts of sports betting Apps filtered by our detectors.

The comparative analysis validates the observation from the previous works and our intuition in designing features discriminating the second download fraud type. It also gives us hints on mitigating such type of download fraud which is discussed in the following section.

\section{RQ3: Stances and Suggestions}
\label{sec:rq3}

There are in total of three parties involved in the attack and defense campaign of download fraud. \textit{App marketers} have the demand in fake downloads, \textit{fraudster agencies} can arrange to inject fake downloads to satisfy such demands, and  app \textit{market operators} suffer from the generated fake downloads. In this section, we first present the stances of three parties probed by us, which help us understand why fake downloads flood in the App markets. Then, we provide suggestions for App market operators on how to mitigate download fraud activities and foster a better App market ecosystem.

\subsection{Stances of Three Parties}

\subsubsection{\textbf{App Marketer}}

For App marketers, we are more interested in why they purchase fake downloads from download agencies instead of promoting their Apps via advertising on the App market. We interview four anonymous App marketers who have purchased the fake downloads. They indicate that not all Apps are eligible to advertise on the App market; purchasing fake downloads has less cost than advertising which only brings a slightly better effect. 

For the crowdturfing fraud, App marketers sometimes use fabricated user acquisition statistics generated by crowdturfing to meet the KPI and cheat the investors. Though legitimate advertisement has a relatively lower cost, it has a more variable conversion rate compared with directly hiring crowd workers to complete elaborate tasks devised by marketers. 

\subsubsection{\textbf{Fraudster Agency}}


Fraudster agencies provide various kinds of services related to App promotion, and most of them are irregular. For the first type of download fraud, fraudster agencies list different prices for different App markets. It suggests that different App markets have different levels of anti-fraud approaches. 


Some fraudster agencies offer an App promotion service bundle called App Store Optimization (ASO). They devise a personalized optimization plan based on the current operation status of an App. ASO is a joint venture between fraudster agencies and App marketers. Such type of service could bring stable cash flow to fraudster agencies and preserve considerable statistics of Apps. We also find some fraudster agencies operate like legitimate IT companies. It illustrates that most of today's download fraud activities become a part of an App's regular promotion tactics.

\subsubsection{\textbf{Market Operator}}


We interview two anonymous operators from non-vendor Android App markets. They claim that the continually evolving fraudsters can easily evade rule-based filtering methods. Fixed thresholds of filtering rules usually result in a high false negative rate, since they only model partial aspects of fraud activities. The market operators only focus on the techniques to filter fraud activities, but seldom think about the way to mitigate fraud activities according to their intentions.

It is worth noting that the interviewees deem that, in some ways, the fake downloads are not 100\% negative. They consider a proper amount of fake downloads would make the App markets thriving and motivate the users' interests in downloading Apps.

\subsection{Suggestions for Market Operators}

According to our previous analysis, we propose five suggestions from A to E on mitigating download fraud for App market operators and other social platforms suffering from similar fraud activities.

\begin{itemize}

    \item \textbf{Adapting the agility of fraudsters.} Due to the continually evolving fraud techniques in the wild, it is better to design an effective and efficient detector, which could detect fraud activities in real time and filter fake downloads immediately. Unsupervised anomaly detection model plus human inspection is a feasible solution in practice. Together with fixing the security breach of the App market system, these approaches will enhance the robustness of the App market ecosystem and reduce its vulnerability as well.
    
    \item \textbf{Building behavior signature database.} Our analysis results in previous sections manifest that most of the fraudsters adopt device ID and IP address resetting techniques during injecting fake downloads. Plus the prevalence of cloud services and IPv6 protocols, IP\&ID blacklisting is no more a valid method to filter bots and fake devices. We also show that rule-based algorithms are vulnerable to attacks. The market operators need to build and update their behavior signature databases, which store the verified suspicious behavior patterns.
    
    \item \textbf{Crafting diversified anti-fraud mechanism.} Our investigation in this work reveals the sophisticated intentions behind download fraud campaigns. Beyond this, there are many external factors causing abnormal traffics. Therefore, the anti-fraud system should consider the motivation of app marketers soliciting fake downloads from multiple views. For different fraud type, a personalized detector would capture more fraud activities and reduce the false positive rate.
    
    \item \textbf{Devising fine-grained advertisement services.} A reasonable App promotion mechanism and advertisement bidding system will attract more App marketers to choose legitimate promotion channels instead of cheating. For example, multi-layered and personalized advertisement pricing would provide more choices for App marketers. Designing better user-advertisement interaction based mechanism could increase the \textit{click-through rate} (CTR) of advertisement and redirect customers from fraudsters to App market advertising system at the same time.

    \item \textbf{Elaborating clear incentives and sanctions.} A strict examination of Apps before they released on App markets could reduce the number of low-quality Apps along with potential fraud activities. Demote the Apps with deceptive activities according to their threat level. It will increase the cost of cheating and thus lower the probability of fraud. It also follows that the proper incentives will divert more app marketers to legitimate promotion channels.

\end{itemize}

\section{Conclusion and Future Work}
\label{sec:discussion}

In this work, we track download fraud activities by setting a honeypot App. We categorize and characterize download fraud activities based on a large scale industrial dataset. With the device vendor flag as ground truth, we further propose and validate several features in identifying bot-generated downloads. By investigating the stances of three parties involved in download fraud, we provide a couple of suggestions for App market operators on how to mitigate download fraud and foster a better ecosystem.

Though vendor flag is considered a valid ground truth to help us dissect the download fraud activities, not all download fraud activities are guaranteed to be found since some advanced download bots may simulate the vendors' devices. Another limitation of our work is that we only investigate download fraud activities in one App market. Some features of the download fraud in other App markets may be different from ours, but the cheating approaches and their intentions should be similar. Moreover, other ensemble classifiers and deep learning models may outperform XGBoost under a new App market, which need to be tested in the future.

We investigate several ways to identify crowdturfing fraud from the server-side log, but could not find clear evidence. However, crowdturfing activities do exist in App markets according to the black market. To identify such crowd workers, a well-annotated dataset with more post-install behavior information needs to be further examined. Fighting fraudsters is a running battle; those selected features may be noneffective after fraudsters camouflage themselves. Thus, devising robust and dynamic countermeasures against fraudsters is another avenue of future research.


\section*{Acknowledgment}

This work is conducted during the internship of Yingtong Dou and Weijian Li in Huawei. 

\bibliographystyle{IEEEtran}
\bibliography{asonam}

\end{document}